\documentclass[doublecol]{epl2} 
\usepackage{amsmath}
\usepackage{amssymb}
\usepackage{textcomp}
\usepackage{graphicx}
\usepackage{grffile}
\usepackage{color}
\usepackage{bm}
\usepackage{tabularx}

\title{Lane formation in a system of dipolar microswimmers}
\shorttitle{Lane-formation of dipolar microswimmers} 

\author{F.~Kogler \and S.~H.~L.~Klapp}
\shortauthor{F. Kogler \etal}

\institute{                    
Institute of Theoretical Physics, Secr. EW 7-1, Technical University Berlin,\\
  Hardenbergstrasse 36, D-10623 Berlin, Germany\\
}


\abstract{
Using Brownian Dynamics (BD) simulations we investigate the non-equilibrium structure formation of a two-dimensional (2D) binary system of dipolar colloids propelling in opposite directions. 
Despite of a pronounced tendency for chain formation, the system displays a transition towards a laned state reminiscent of lane formation in systems with isotropic repulsive interactions.
However, the  anisotropic dipolar interactions induce novel features: First, the lanes have themselves a complex internal structure characterized by chains or clusters.
Second, laning occurs only in a window of interaction strengths. We interprete our findings by a phase separation process and simple force balance arguments.}

\begin{document}

\maketitle

\section{Introduction}
Lane formation is a protoype of a non-equilibrium self-organization process, 
where an originally homogenous mixture of particles (or other types of "agents") moving in opposite directions
segregates into macroscopic lanes composed of different species. This ubiquitous phenomenon occurs, e.g., in driven binary mixtures
of colloidal particles~\cite{leunissen2005,loewen:exp,loewen:first}
and migrating macroions~\cite{conduction:lane}, in binary plasmas~\cite{plasma:lane2,Suetterlin2009}, 
but also in "self-propelling" systems with aligned velocities such as bacteria in channels~\cite{lane:unidirectional} and humans in pedestrian zones~\cite{Helbing2000}. 
In particular, studies of charged colloids have revealed many fundamental aspects of laning such as the impact of density~\cite{loewen:DFT-phasediagram}, the role of hydrodyamics~\cite{loewen:hydro}, 
the accompanying microscopic dynamics (particularly, the so-called dynamical locking)~\cite{loewen:exp}, and the impact of anisotropic friction~\cite{attraction:lane}.

All of these models involve {\em isotropic} interactions between the colloidal particles. However,
recent progress in colloidal chemistry has generated a variety of complex, anisotropic particles (see, e.g.,~\cite{review:janus1,review:janus2,shiftdip:exp2}),
which can perform controlled translational motion in an external field~\cite{gangwal:active,gangwal:nets}, like simple charged colloids, 
but display complex self-assembly behaviour already in equilibrium~\cite{gangwal:nets,schmidle:nets}.
In driven ensembles of such particles one may therefore expect a wealth of new phenomena induced by the
interplay between self-assembly, on the one hand, and dynamical self-organization processes such as laning, on the other hand. 
The consequences are so far, only poorly understood, contrary to the widely studied case of active (self-propelled) particles (~\cite{nonequi:review1, Bechinger2013, Redner2013_1}).

In the present letter we discuss, based on particle-based Brownian Dynamics (BD) simulations,
a prototype of field-propelled complex colloids, that is, spheres with dipolar interactions. Our model is inspired by 
real systems of metallodielectric "Janus" spheres with two dielectric parts acquiring different (induced) dipole moments in an external electric field~\cite{gangwal:passive}. 
Previous experimental studies in a quasi-two dimensional (2D) setup with an in-plane field
have shown that these particles perform straight motion {\em perpendicular} to the field~\cite{gangwal:active}, with two possible directions depending on the orientation of the hemispheres. 
Ensembles of such particles therefore should exhibit lane formation, with the additional feature of strongly anisotropic interactions favoring the formation of staggered chains. 

Based on a suitable model system, we find indeed several new phenomena: First, despite a pronounced tendency to aggregate into chains
perpendicular to the driving force, the dipolar interactions can induce laning at densities where purely repulsive
systems are mixed. At larger interaction strengths, however, we observe a breakdown of lane formation. Moreover, by comparing
the lane formation with that in a simpler system governed by isotropic, attractive Lennard-Jones (LJ) interactions, we show that the main 
laning mechanism is, in fact, the effective (angle-averaged) {\em attraction} between the dipoles. Indeed, for both systems the onset of laning roughly occurs at coupling strengths related to 
{\em equilibrium phase separation}. Finally, lanes disappear when the thermal energy exceeds the attraction (corrected by the drive) between two particles.
\section{Theoretical models and methods}
\begin{figure}
  \centering
  \includegraphics[width=0.45\textwidth]{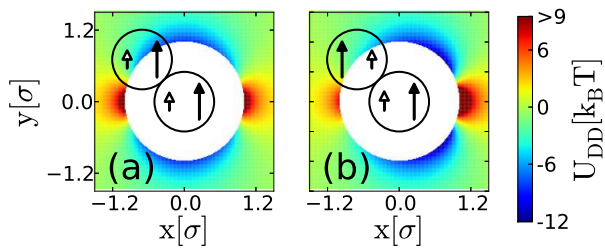}
  \caption{\label{fig:1} (Color online) Potential energy 'landscape' due to the total dipolar interaction (see color code) 
 for two DDPs of diameter $\sigma$ (solid circles) with (a) $s_i=s_j=-1$ and (b) $s_i=1,s_j=-1$ in a top-view perspective for $(\mu^{1})^*=1.58$. 
 The particle dipole moments are indicated by arrows with a white [black] head for 
 $\boldsymbol{\mu}^{(1)}$ [$\boldsymbol{\mu}^{(2)}$]. The white circle around the centered particle indicates the excluded area. }
\end{figure}
The experimental particles motivating this work are gold-patched dielectric spheres confined between two glass plates, forming a quasi-2D geometry \cite{gangwal:active}.
Application of an in-plane AC electric field aligns the plane between gold patched and dielectric part along the field and induces dipole moments in both, 
the gold and the dielectric part of the particles, with the gold's dipole being significantly larger due to its larger polarisability. 
At low frequencies these two induced dipoles have the same orientation (along the field). Placed into an aqueous solution, one observes~\cite{gangwal:active} 
a spontaneous motion of each particle in the direction {\em orthogonal} to the field and away from the particle's gold patch, the underlying mechanism being asymmetric flow of solvent charges
(induced-charge electrophoresis)~\cite{gangwal:active,ICEP:details}.

To examine the collective behaviour of these driven particles, we perform BD simulations in a 2D quadratic cell of size $L^2$ with periodic boundary conditions. 
The cell contains $N=800$ spherical particles with equal diameter $\sigma$ defining a unit length. Particle positions are denoted by $\boldsymbol{r}_{i}=x_{i}\boldsymbol{e}_{x}+y_{i}\boldsymbol{e}_y$ ($i=1,..,N$) 
with the unit vectors $\boldsymbol{e}_{x}$ and $\boldsymbol{e}_{y}$ defining a 2D coordinate frame. The external, electric field $\boldsymbol E_{ext}$ points along the $y$-axis.
Sterical interactions are modeled by a soft-sphere (SS) potential $U_{SS}(\boldsymbol r_{ij}) = 4\epsilon_{SS}\left((\sigma/r_{ij})^{12} - (\sigma/r_{ij})^{6}+1/4\right)$ 
truncated at $r^c_{ij} = 2^\frac{1}{6}\sigma$, where $r_{ij}=|\boldsymbol r_{j}-\boldsymbol r_{i}|$ is the particle distance. The strength of repulsion is set to $\epsilon^*_{SS}=\epsilon_{SS}/k_BT=10$ 
where $k_B$ is Boltzmann's constant and $T$ is the temperature.
To mimic the impact of the propulsion force we randomly assign to each particle a fixed vector $\boldsymbol s_i=s_i\boldsymbol e_x$ 
which points either to the 'right' ($s=1$) or to the 'left' ($s=-1$) with probability $0.5$, respectively. 
Thereby a random fifty-fifty mixture of two different particle 'species' is created. Each particle is then subject to a 
constant but orientation-dependent driving force $\boldsymbol f_{d,i}=f_d\boldsymbol s_i$. Changes of particle orientation ($s_i$) are not considered here.

To incorporate the dipolar interactions we assume that each particle $i$ bears two point dipole moments $\boldsymbol\mu^{(1)}_{i}$ and $\boldsymbol\mu^{(2)}_{i}$ whose 
orientation is fixed along the external field, i.e. $\boldsymbol{\mu}_i^{(\alpha)}=\mu_i^{(\alpha)}\boldsymbol{e}_y$, $\alpha=1,2$. 
Moreover, the two dipole-moments are shifted out of the particle center 
by $\boldsymbol\delta^{(1)}_{s_i}=\delta\boldsymbol s_i$ and $\boldsymbol\delta^{(2)}_{s_i}=-\delta\boldsymbol s_i$ with $\delta=0.25\sigma$. The values of $\mu_i^{(\alpha)}$ are hold constant 
(that is, we neglect variations of the local field) and set differently in order to mimic the strong asymmetry of the metallo-dielectric janus-particles. Specifically, we choose $\mu_i^{(2)} = 2\mu_i^{(1)}$ and 
set the dimensionless dipole strength to $\mu^{*}\equiv{\mu_i^{(2)}}^*=\mu_i^{(2)}/\sqrt{\sigma^3k_BT}$. Test simulations revealed that the system's behaviour is only weakly sensitive to the ratio 
$\mu_2 / \mu_1$, whereas changing the shift parameter $\delta$ has a strong effect. Here we stick to a value of $\delta$ already used in Ref.~\cite{schmidle:nets} for the case of oppositely oriented dipoles.
The distance between two arbitrary dipole moments $\boldsymbol\mu^{(\alpha)}_{i}$ and $\boldsymbol\mu^{(\gamma)}_{j}$ of 
different particles is given by $\boldsymbol r_{s_is_j}^{\alpha\gamma} = \boldsymbol r_{j} + \boldsymbol\delta^\gamma_{s_j} - (\boldsymbol r_{i} + \boldsymbol\delta^\alpha_{s_i} ) $.
The dipolar interaction between the two of them is calculated via the 3D point dipole potential 
$U_{dip}(\boldsymbol r_{ij}, \boldsymbol\mu_i^{(\alpha)},\boldsymbol\mu_j^{(\gamma)},s_i,s_j)=\boldsymbol \mu^{(\alpha)}_{i}\cdot\boldsymbol \mu^{(\gamma)}_{j} (r_{s_is_j}^{\alpha\gamma})^{-3}-3(\boldsymbol\mu^{(\alpha)}_{i}\cdot\boldsymbol r_{s_is_j}^{\alpha\gamma})(\boldsymbol\mu^{(\gamma)}_{j}\cdot\boldsymbol r_{s_is_j}^{\alpha\gamma})(r_{s_is_j}^{\alpha\gamma})^{-5}$, 
and the total dipolar interaction between two particles is given by the sum $U_{DD}(\boldsymbol r_{ij},s_i,s_j) = \sum^2_{\alpha,\gamma = 1} U_{dip}(\boldsymbol r_{ij}, \boldsymbol\mu_i^{(\alpha)},\boldsymbol\mu_j^{(\gamma)},s_i,s_j)$. 
We use the standard 2D Ewald summation method to handle the long-range character of the dipole interactions~\cite{martial:ewald}.
Combining $U_{DD}$ and the formerly introduced SS potential defines the 'double-dipole particle' (DDP) model $U_{DDP}=U_{DD}+U_{SS}$.
The angle dependency for distances $r_{ij}>\sigma$ is illustrated in fig.~\ref{fig:1}. 
It is seen that two identical DDPs ($s_i=s_j$) prefer a head-to-tail configuration similar to "single-dipole" particles with one dipole in their center. 
In contrast, for DDPs with $s_i \neq s_j$, the most attractive configurations 
are staggered in the sense that one DDP is shifted towards the gold-patched  'back' of the other particle. The corresponding minimum energy $U^{min}_{DD}(s_i\neq s_j)$ 
is roughly $4k_BT$ larger than in the head-to-tail condiguration $U^{min}_{DD}(s_i=s_j)$. 
Finally, for reasons discussed later we also consider a driven LJ fluid characterized by the pair 
potential $U_{LJ} = 4\epsilon((\sigma / r_{ij})^{12} - (\sigma / r_{ij})^{6}$ with cut-off at $2.5\sigma$ and dimensionless interaction strength $\epsilon^*=\epsilon/k_BT$.
For all model systems the overdamped BD equations of motion are given by
$\gamma \dot {\boldsymbol r}_{i} = \sum^N_{j=1} \boldsymbol \nabla U_X(ij) + \boldsymbol f^s_{d,i} + \boldsymbol \zeta_i$
where $ \gamma$ is the friction constant, 
$U_X(ij)$ is a specific pair potential (with $X=SS$, $DDP$, or $LJ$), and $\boldsymbol \zeta_i$ is a Gaussian noise vector which acts on particle $i$ and fulfills the relation
$\langle \boldsymbol \zeta_i \rangle = 0$ and $\langle \boldsymbol \zeta_i (t) \boldsymbol \zeta_j (t') \rangle = 2 \gamma k_BT \delta_{ij}\delta(t-t')$. Hydrodynamic interactions (HI) are neglected,
in accordance with earlier studies~\cite{loewen:hydro} revealing that HI do not alter lane formation qualitatively.
The BD equations are solved via the Euler scheme~\cite{Ermak1975} with an integration stepwidth $\Delta t = 10^{-5}\tau_b$, where $\tau_b$ is a brownian timescale defined by $\tau_b=\sigma^2 \gamma/ k_BT$.

\section{Target quantities} To quantify the degree of laning we employ the laning order parameter $\Phi$ defined in \cite{loewen:first}. Perfectly laned states  correspond to $\Phi=1$. 
However, the system can visually appear as being laned already at much smaller values of $\Phi$ (e.g., $\Phi\approx0.2$), especially at high densities. To characterize the local structure 
in $x$-direction we calculate the radial distribution function between particles of the same type
\begin{equation}
\label{eq.0}
g^s(x) = \frac{2}{\rho N}\langle\sum^N_{i=1}\sum^N_{j\neq i}\delta(x-|x_{ij}|)\Theta(\sigma-|y_{ij}|)\Theta(s_is_j)\rangle
\end{equation}
with $x_{ij}$ ($y_{ij}$) being the $x$ ($y$)-component of $\boldsymbol{r}_{ij}$, and $\delta$ and $\Theta$ are the delta- and the Heavyside step function, respectively. 
By normalization, $g^s(x)$ decays to $2$ for $x \to \infty$ in a single-species fluid and to 1 in a completly mixed binary fluid. By interchanging $x$ and $y$ 
in eq.~(\ref{eq.0}) we additionally define the correlation function $g^s(y)$ in $y$-direction. 
\section{Results}
\begin{figure}
  \centering
  \includegraphics[width=0.49\textwidth]{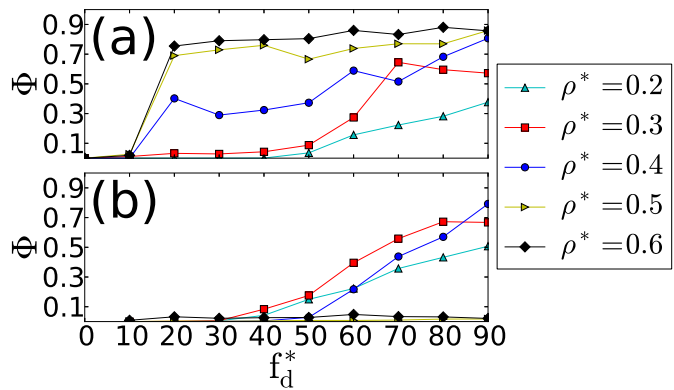}
  \caption{(Color online) Laning order parameter as function of driving force for (a) the DDP system ($\mu^{*}=1.58$) and (b) the SS system ($\epsilon_{SS}^{*}=10k_BT$) at different densities.}
  \label{fig:2}
\end{figure}
\begin{figure}
  \centering
  \includegraphics[width=0.49\textwidth]{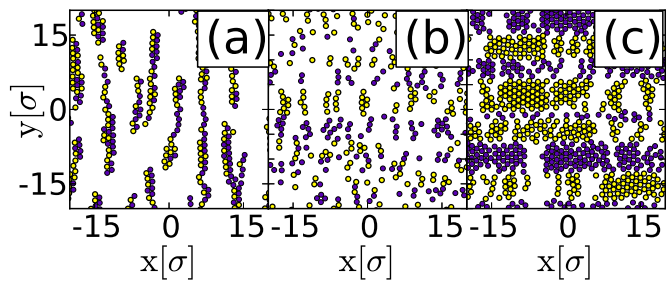}
  \caption{(Color online) Snapshots of the DDP system at (a) $f^*_d=0$, $\rho^*=0.2$, (b) $f^*_d=80$, $\rho^*=0.2$ ($\Phi \approx 0.2$) and (c) $f^*_d=80$, $\rho^*=0.5$ ($\Phi \approx 0.75$). 
 Dark (bright) particles: $s=1$ ($-1$), $\mu^{*}=1.58$.}
  \label{fig:3}
\end{figure}
\begin{figure}
 \includegraphics[width=0.45\textwidth]{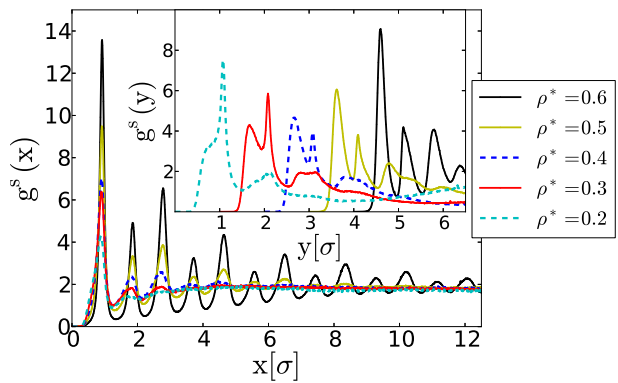}
  \centering
  \caption{(Color online) Pair correlation functions between particles of the same species along the $x$-direction (main plot) and the $y$-direction (inset). The data pertain to $f_d^*=80$ and different densities. 
  In the inset, curves for different densities have been shifted by one $\sigma$ to enhance visibility. \label{fig:4}}
\end{figure}
In fig.~\ref{fig:2}(a) we plot the laning order parameter $\Phi$ of DDP systems for different densities $\rho^{*}=\rho\sigma^2$ as functions of the dimensionless driving force 
$f_d^{*}=f_d\sigma/k_B T$.
At low driving forces ($f^*_d \leq 10$) the order parameter is essentially zero for all densities considered, reflecting the absence of lanes.
The corresponding local structure at $f^*_d=0$ and $\rho^{*}=0.2$ is illustrated by the simulation snapshot shown in fig.~\ref{fig:3}(a). 
One observes the formation of 'staggered' chains, consistent with the most attractive
pair configurations illustrated in fig.~\ref{fig:1}(b), and in qualitative agreement with experiments of metallodielectric particles under field conditions where self-propulsion is absent~\cite{gangwal:passive}. 
Increasing the driving force from zero ($f_d^*\lesssim 10$), these staggered chains are pushed against each other and form one large aggregate consisting of both species.
Upon further increase of $f^*_d$, all of the systems considered in fig.~\ref{fig:2}(a) display laned 'states' characterised by 
non-negligible values of $\Phi$. For small densities ($\rho^{*} < 0.4$) this lane formation occurs gradually and significant values of $\Phi$ are reached
only at high driving forces $f^*_d > 50$. In contrast, dense systems display a rather steep laning "transition" 
at a driving force $f^*_d \approx15$. A visualisation of exemplary laned states at $f^*_d=80$ and two densities is given in fig.~\ref{fig:3}(b) and (c).

To elucidate the impact of dipolar forces on the functions $\Phi(f^*_d)$ we plot in fig.~\ref{fig:2}(b) corresponding data for the driven SS system. 
At low densities $\rho^*\leq0.4$ the systems behave similar to their DDP counterparts. However, at high densities ($\rho^*>0.4$) 
there is {\em no} lane formation in the (purely repulsive) SS system, in striking contrast to the behaviour of the DDP systems. We will come back to this point below.
Here we first consider the internal structure of the lanes in the driven DDP systems. Visual inspection of figs.~\ref{fig:3}(b) and (c) suggests 
that lanes consist of clusters of particles of the same type. 
At low densities (see fig.~\ref{fig:3}(a)) we find mostly short chains oriented in $y$-direction, 
which reminds of the equilibrium structure of single dipole particles in external fields \cite{weis:quasi-dipoles}. 
Clearly, such chains are absent in simpler (isotropic) lane-forming systems.
At higher densities (see fig.~\ref{fig:3}(b)) the particles tend to form clusters along the driving force.
Their internal structure somewhat resembles that in a (hexagonal) crystal; in fact, the corresponding translational order parameter~\cite{Dzubiella2002_crystal} is larger than that observed at low $f_d^{*}$.
For additional quantitative characterization we consider the correlation functions $g^s(y)$ and $g^s(x)$ plotted in fig.~\ref{fig:4}. 
For the lowest density considered ($\rho^*=0.2$) $g^s(y)$ reveals a pronounced peak at $y=1\sigma$ and a second, 
smaller one at $y=2\sigma$, indicating chains of up to three particles along the field direction. Furthermore, 
we observe a 'hump' appearing directly before the first peak. This 'hump' results from bended and perturbed chains 
as well as from a few clusters elongated in $x$-direction (see fig.~\ref{fig:3}(b)). Increasing the density the hump 
transforms into a pronounced peak, which finally overtakes the peak at $y=1\sigma$ at $\rho^*=0.4$. 
At this density, chains in $y$-direction have lost their dominance as structural elements. 
Instead, clusters spread out in $x$-direction. This interpretation is supported by
the function $g^s(x)$ which displays a similar qualitative change at $\rho^*=0.4$. 
We note that these structural details are sensitive to the shift parameter $\delta$ characterizing 
the dipole positions inside the particles: Indeed, as revealed by test simulations, 
smaller values of $\delta$ tend to surpress the lane formation in the "chaining regime" and enhance the occurence of chains in the "cluster regime". 

\section{Mechanisms of lane formation}
\begin{figure}
  \centering
  \includegraphics[width=0.35\textwidth]{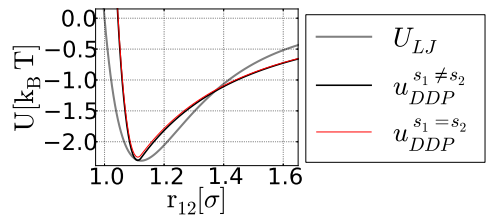}
  \caption{(Color online) Angle-averaged DDP potential for $s_1 = s_2$ and $s_1 \neq s_2$ at $\mu^*=1.58$. Included is the LJ potential at $\epsilon^{*}=2.3$. 
  }\label{fig:5}
\end{figure}

\begin{figure*}
  \centering
  \includegraphics[width=0.3\textwidth]{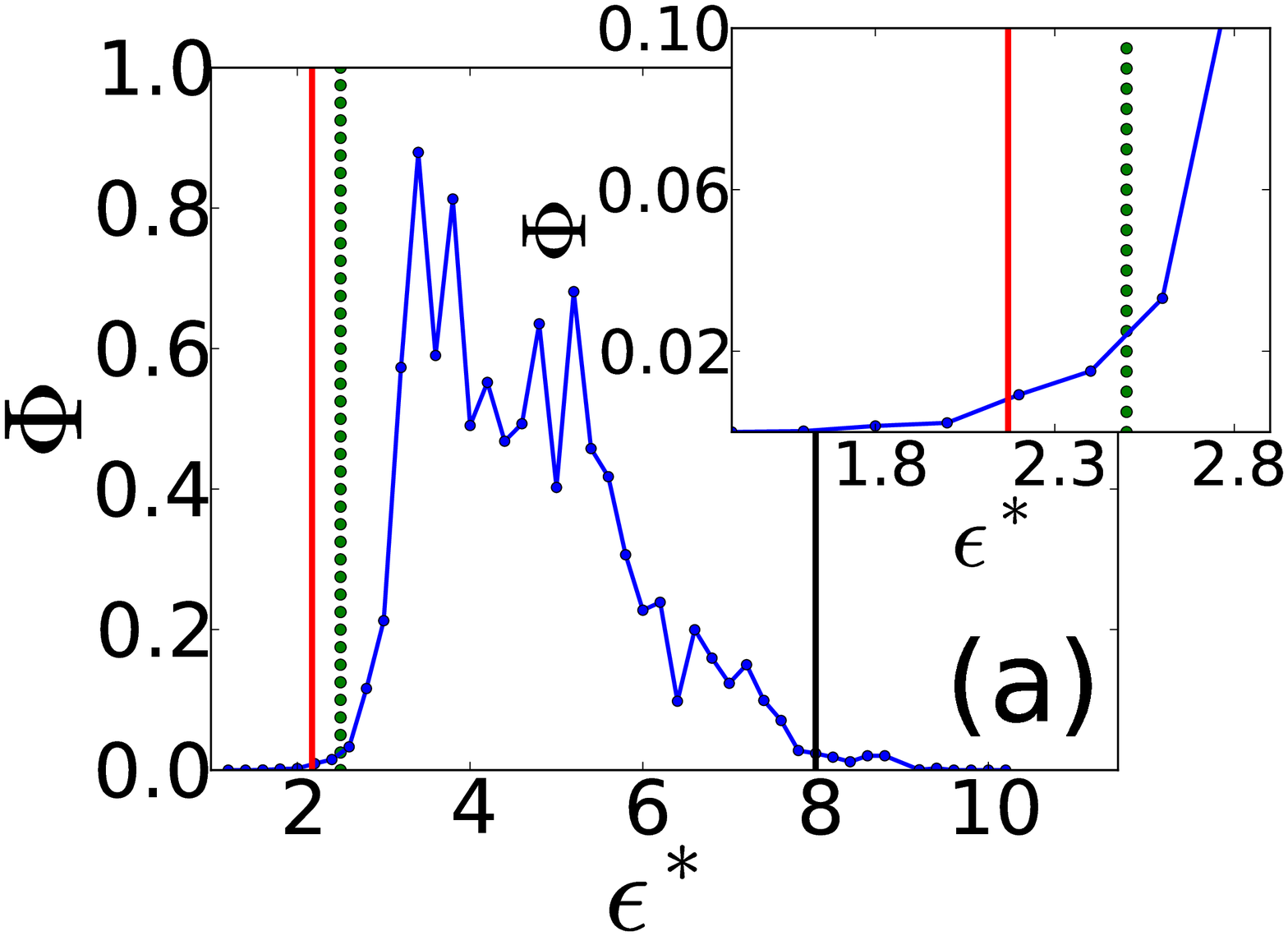} \includegraphics[width=0.3\textwidth]{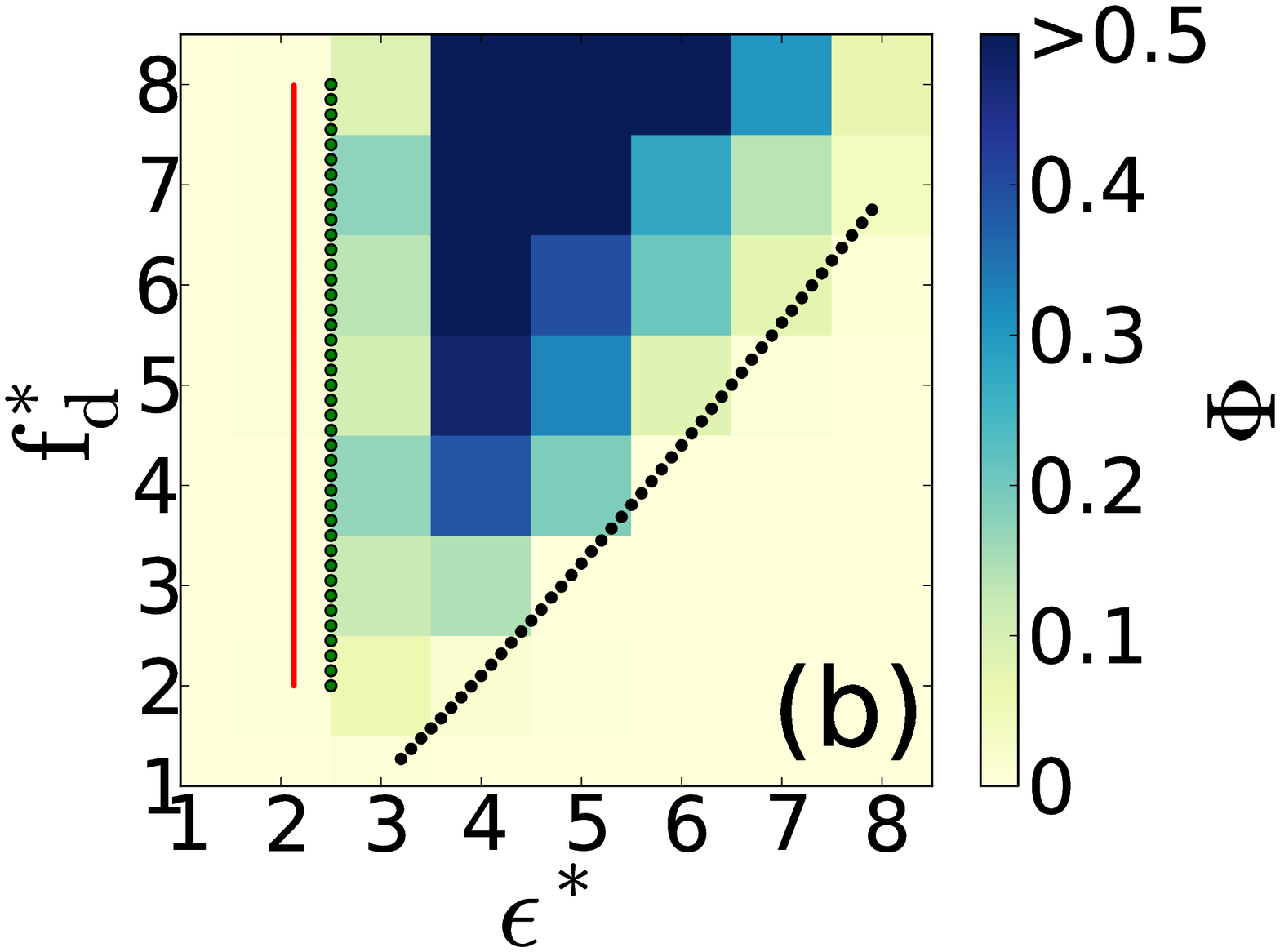} \includegraphics[width=0.3\textwidth]{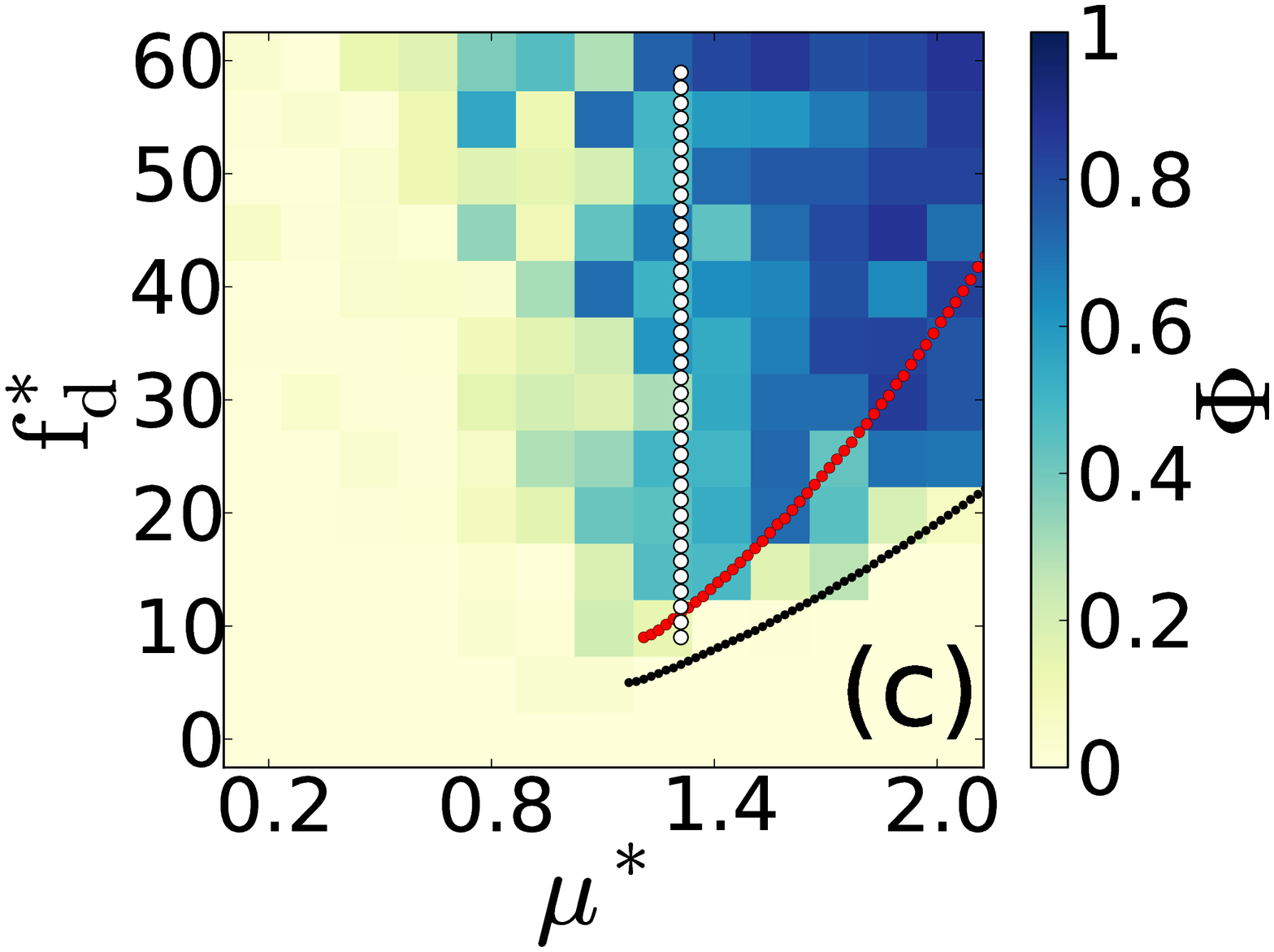}
  \caption{(Color online) (a) Laning parameter versus coupling strength for the LJ system at $f^*_d=7$, $\rho^*=0.5$. Inset: Enlarged view for small $\epsilon^{*}$. 
(b) Laning "state diagram" of the LJ system. In (a) and (b), 
 red, dotted green and black lines correspond to $\epsilon^*_c$, $\epsilon^*_t$, and $\epsilon^{*}_{max}$, respectively. 
(c) Corresponding state diagram of the DDP system.  The dotted white, red(black) lines correspond to $\mu^{*}_s$  and $\mu^{*}_{max}$ $at$ $y_0 = 0.645(0.9)\sigma$ .\label{fig:6}}
\end{figure*}

In the following we concentrate on the question {\em why} the dipolar interactions strongly enhance lane formation compared to 
the purely repulsive SS model, as revealed by the order parameter plots in fig.~\ref{fig:2}. We concentrate on the density $\rho^*=0.5$, where the differences are particularly pronounced.
Besides the anisotropy, another main feature characterizing the DDP interactions is that they are {\em effectively attractive}. This is seen from the angle-averaged potential 
\begin{equation}
\label{eq.1}
u_{DD}(r_{12})=\frac{1}{2\pi}\int_0^{2\pi} U_{DD}\;(\boldsymbol r_{12},s_1,s_2)\; d\Omega_{12},
\end{equation}
with $\Omega_{12}$ being the angle of the connecting vector $\boldsymbol{r}_{12}$ relative to the $x$-axis.
Numerical data for the cases $s_1=s_2$ and $s_1\neq s_2$ are shown in fig.~\ref{fig:5}.
Clearly both angle-averaged potentials display a pronounced, attractive potential well with nearly indentical depths. 
In fact, the angle-averaged potential somewhat resembles that of an LJ fluid (also plotted in fig.~\ref{fig:5}), 
disregarding differences in the width of the attractive well and the decay of the potentials with $r_{12}$.
To understand the role of the {\em effective} attraction in our DDP system on the lane formation we therefore consider, 
as a first step, the corresponding behaviour of an LJ fluid. For simplicity, we choose the same attraction strength $\epsilon^*$ for 
both, different and same species. Results for the order parameter $\Phi$ of a driven LJ system as function of $\epsilon^{*}$ at fixed $f_d^*$ are shown in fig.~\ref{fig:6}(a). 
For very small values of $\epsilon^{*}$ the order parameter is negligible, consistent with the behaviour of the pure SS system in fig.~\ref{fig:2}(b). 
Upon increase of $\epsilon^{*}$, lanes first appear at a certain value 
and then {\em disappear} again at a substantially larger coupling. A similar {\em reentrance} of the non-laned (mixed) state 
occurs at other driving 
forces, as seen from the laning "state diagram" (for $\rho^*=0.5$) presented in fig.~\ref{fig:6}(b). 
The state diagram moreover reveals that the {\em onset} of lane formation occurs at a coupling strength of $\epsilon^{*}\approx 2.5$ 
(see dotted green lines in figs.~\ref{fig:6}(a)-(b)) quite independent of $f_d^*$ (taking the value $\Phi =0.02$ as a lower limit for laning), while the 
coupling related to the disappearance of lanes appears to be a {\em linear} function of $f_d^{*}$. 
Given the lower boundary $\epsilon^{*}\approx 2.5$, it is interesting to make a connection to the {\em equilibrium} phase diagram of the 2D LJ fluid: 
This system has a critical point (gas-liquid condensation) at $\rho^*_c=0.35$ and $T^{*}_c\approx 0.46$, corresponding to a critical 
coupling strength $\epsilon^{*}_c\approx 2.17$~\cite{triple:LJ,smit_frenkel:LJ2D}. The triple point (gas-liquid-solid) coupling 
strength is given as $\epsilon^{*}_t\approx 2.5$ (with corresponding liquid density $\rho_t^{liquid} \gtrsim 0.6$~\cite{triple:LJ}). From this we can conclude 
that the non-driven LJ system at $\rho^*=0.5$ and $\epsilon^{*}=2.5$ is in a strongly correlated state which is, in fact, 
thermodynamically unstable, i.e., it lies {\em within} the coexistence curve of the gas-liquid transition. 
Moreover, revisiting again fig.~\ref{fig:6}(a)-(b) we see that $\epsilon_c$ provides a lower limit of laning.
These observations suggest that the equilibrium phase separation is a {\em prerequisite} for lane formation in the driven LJ system.

We now turn to the breakdown of laning at large coupling strengths.
Here, the external force driving two unlike particles ($s_i\neq s_j$) away from one another competes with the attractive LJ forces.
To quantify this competition we construct an {\em effective} pair interaction $U^{LJ}_{eff}(x_{ij})$ between unlike particles, which are initially in contact
and then driven apart in $x$-direction.
To account for different vertical positions $y_{ij}$, we first average the $x$-component of the LJ force $F_{LJ,x}=-\partial_{x_i} U_{LJ}$, over the angle
$\phi$ between $\boldsymbol{r}_{ij}$ and $\boldsymbol{e}_y$ (at fixed distance) yielding
$\bar{F}_{LJ,x}=\pi^{-1}\int_0^\pi \sin\phi F_{LJ}d \phi=
2F_{LJ}/\pi$ with $F_{LJ}=-\partial_r U^{LJ}$. Integration (setting now $y_{ij}=0$) yields the potential $\bar{U}_{LJ}(x_{ij})= 2/\pi U_{LJ}(x_{ij})$. Adding the
linear potential associated to the external force yields $U^{LJ}_{eff}(x_{ij})=\bar{U}_{LJ}(x_{ij}) - f_d x_{ij}$. 
For small driving forces $U^{LJ}_{eff}$ displays a local minimum $U_{min}$ and maximum $U_{max}$, and thus a potential barrier 
$\Delta U=|U_{max}-U_{min}|$. 
Unlike particles tend to stick together (against the drive) if the mean kinetic energy per 
particle $2 \times k_BT/2$ is smaller than $\Delta U(f_d,\epsilon)$. Solving
numerically the equation $\Delta U = k_BT$ for various $f_d^{*}$ 
we obtain values $\epsilon^{*}_{max}(f_d^{*})$ indicated by the dotted black line
in fig.~\ref{fig:6}(b). Clearly, our simplified model describes
the breakdown of lane formation very well. Moreover, having introduced $U^{LJ}_{eff}$, we can also understand the fact that the onset of laning occurs at coupling strength slightly larger than the critical coupling in equilibrium, $\epsilon_c$. The driving force effectively reduces the attraction between unlike particles, yielding a shift of phase separation.

Given this background, we now consider in fig.~\ref{fig:6}(c) the state diagram of the full, anisotropic DDP fluid ($\rho^*=0.5$) in the parameter plane spanned by driving force
and coupling strength. The latter is now given by $\mu^*$ (the soft-sphere repulsion is fixed).
Clearly, the structure of the state diagram resembles that of the LJ system. In particular, the onset of laning depends only weakly on $f_d^*$, and there is an upper boundary for $\mu^*$ beyond which laning disappears.

This prompts the question whether the onset of laning in the driven DDP system can be related to an equilibrium fluid-fluid phase separation, as in the LJ system. 
We note that already the equilibrium DDP system is a true binary mixture, since the full pair interactions depend on $s_i$, see fig.~\ref{fig:1}. Therefore the possible fluid-fluid transitions (if existent at all) are, in general, combinations of condensation and demixing.
In the present study we did not carry out simulations to explore these questions properly. Still, one can perform some estimates based on mean-field density functional theory (DFT). 
We focus on the occurence of condensation (rather than demixing), because the angle-averaged (i.e. mean-field) interaction between unlike DDPs is nearly the same as that between like particles (see fig.~\ref{fig:5}).
The corresponding stability condition of the homogeneous, mixed phase is that the isothermal compressibility, $\chi_T$, has to be positive.
According to Kirkwood-Buff theory~\cite{Kirkwood_Buff} one has (for a symmetrix binary mixture composed of species $A$ and $B$) 
$\chi_T^{-1}\propto 1-(\rho/2)\left(\tilde c_{AA}(0)+\tilde c_{AB}(0)\right)$, where $\tilde{c}_{AA(AB)}(0)$ 
are the Fourier transforms of the direct correlation functions (DCFs) $c_{AA(AB)}(\boldsymbol{r}_{12})$
in the limit of long-wavelengths ($k\rightarrow 0$). In our case, $AA$ ($AB)$ corresponds to $s_1=s_2$ ($s_1 =-s_2$). Furthermore, we approximate
the DCFs according to a random phase approximation, that is,
$c_{s_1 s_2}(\boldsymbol{r}_{12})=c_{HS}(r_{12})\theta(\sigma-r_{12})-(k_BT)^{-1}U_{DD}(\boldsymbol{r}_{12},s_1,s_2)\theta(r_{12}-\sigma)$, where
$c_{HS}$ is the Percus-Yevick DCF of a pure hard-sphere fluid~\cite{DCF}. 
The Fourier transforms of the second, mean-field like contribution to the DCFs yield essentially the spatial integral over the effective potentials
defined in eq.~(\ref{eq.1}).
Numerical investigation of the resulting expression for $\chi_T$ at $\rho^{*}=0.5$ reveals that, upon increasing $\mu^{*}$ from zero (where
$\chi_T$ reduces to the hard-sphere compressibility), $\chi_T$ becomes indeed negative at $\mu^{*}_s=1.31$, indicating an instability ("spinodal point") related to condensation.  
Thus, simple mean-field DFT predicts the existence of a condensation transition in the equilibrium DDP fluid. Further, for $\rho^{*}=0.5$, the transition should occur at
a coexistence value $\mu^{*}_{coex}\lesssim \mu^{*}_s$; its actual value is, however, more eloborate to determine~\cite{footnote1}. BD test simulations of the non-driven system at $\rho^{*}=0.5$ and dipole moments larger than $\mu^{*}_s$ reveal indeed a phase separated structure consisting
of thick columns (involving both particle species) and large voids in between.

The question now is, does $\mu^{*}_s$ play a decisive role
for the {\em driven} DDP system?
Considering fig.~\ref{fig:6}(c) we find that $\mu^{*}_s$ (indicated by a white line) yields indeed a good estimate for the onset of laning at low values of the driving force ($f_d^{*}\lesssim 30$). 
This suggests that, similar to the LJ system, equilibrium phase separation is a prerequisite of laning at the density considered. 
Only for larger driving forces ($f_d^{*}\gtrsim 30$) the estimate worsens (this effect was not observed
in the LJ system where, however, calculations where restricted to small $f_d^{*}$).

The breakdown of laning at large $\mu^*$ can be estimated, similar to the LJ system,
by constructing an one-dimensional ($x$-dependent) potential, in which 
the DDP interaction between unlike particles competes with the driving force.
We consider examplary configurations $y_{ij}=y_0$ (a simple force average over different 
configurations $y_{ij}$ is not appropriate due to the interaction anisotropy)
and obtain the potential $U^{DDP}_{eff}(x_{ij}) = U_{DDP}(x;y_0) - f_d x_{ij}$ which displays a barrier $\Delta U(y_0,f_d,\mu)$.
Figure~\ref{fig:6}(c) includes the numerical solution $\mu^{*}_{max}(f^*_d)$ (dotted red line)
of the equation $\Delta U=k_B T$ at $y_0=0.645\sigma$, corresponding to the most attractive,
"staggered" configuration at contact (see Fig.~\ref{fig:1}(b)). Clearly, this line describes the breakdown
qualitatively, but overestimates the influence of attraction. 
Quantitatively better results are obtained with $y_0=0.9\sigma$ (dotted black line in fig.~\ref{fig:6}(c)), corresponding to a less attractive initial configuration. 
The "true" potential should be seen as a (weighted) average over different $y_0$. 
Finally, the reduction of attraction expressed in the effective potential explains
the shift of the onset of laning with larger $f^*_d$. 

\section{Conclusions}
Our study reveals new und unexpected aspects of laning in driven colloidal mixtures with anisotropic, dipolar interactions between the particles. Most importantly,
anisotropic interactions do not surpress the lane formation despite 
the strong tendency of the particles to self-assemble into chains (which, in the present model, are oriented {\em perpendicular} to the force). 
Rather, we observe {\em stabilization} of laning (relative to repulsive systems) in a window
of interaction strength, with the anisotropic interactions influencing strongly the structure of the lanes.
This stabilization is due to the effective, angle-averaged attraction inherent in our model, which eventually leads to equilibrium phase separation (condensation). Interestingly,
the corresponding coupling strengths also give a good estimate for the {\em onset} of laning upon at fixed driving force, and we have observed
similar behaviour in the simpler (isotropic) LJ system. This indicates that laning in systems with attractive interactions is intimately related to phase separation, irrespective
of their detailed form. Our observations should be directly measurable in real systems of field-propelled metallodielectric particles~\cite{gangwal:active}. 
Moreover, we expect our findings to be transferable
to other driven colloidal systems with direction-dependent interactions, examples being "patchy" particles which display both 
equilibrium aggregation {\it and} condensation~\cite{Sciortino2006}, or dipolar colloids under shear flow~\cite{dipshear}.

Starting from the present study, there is several further intriguing questions to be explored. For example, is the behaviour observed here related to a Raleigh-Taylor (RT) instability occuring in
driven, macroscopically phase-separated mixtures?  In the context of RT, the critical wavelength~\cite{Loewen:Raleigh_Taylor} 
can be estimated by a simple criterion involving the surface tension; this might provide a way to quantify 
the lane width of our system.Indeed, the present simulation data are not yet conclusive to predict, e.g., a relation between lane width and dipolar coupling. 
We also point out possible connections to driven,phase-separating magnetic mixtures~\cite{rat:active}.
Finally it would be worth to explore connections to active fluids consisting of self-propelled agents with attractive interactions. Indeed, the interplay of
self-organization in such systems and equilibrium phase separation is currently a very lively field of research~\cite{Schwarz-Linek2012,Redner2013_1,Redner2013_2, Bechinger2013,bialke2015}.

We gratefully acknowledge financial support of DFG via the IRTG 1524.

\end{document}